# Resistive Switching Phenomena of HfO$_2$ Films Grown by MOCVD for Resistive Switching Memory Devices

**Min Ju Yun, Sungho Kim\*, and Hee-Dong Kim\*\***

*Department of Electrical Engineering, Sejong University, Seoul 143-747*

The resistive switching phenomena of HfO$_2$ films grown by metalorganic chemical vapor deposition was studied for the application of ReRAM devices. In the fabricated Pt/HfO$_2$/TiN memory cells, the bipolar resistive switching characteristics were observed, and the set and reset states were measured to be as low as 7 µA and 4 µA, respectively, at $V_{READ} = 1$ V. Regarding the resistive switching performance, the stable RS performance was observed under 40 repetitive dc cycling test with the small variations of set/reset voltages and currents, and good retention characteristics over $10^5$ s in both LRS and HRS. These results show the possibility of MOCVD grown HfO$_2$ films as a promising resistive switching materials for ReRAM applications.

\*E-mail: sungho85.kim@sejong.ac.kr, tel: +82-2-6935-2422, fax: +82-2-3408-4329

\*\*E-mail: khd0708@sejong.ac.kr, tel: +82-2-6935-2423, fax: +82-2-3408-4329

## I. INTRODUCTION

Resistive random access memory (ReRAM) has recently attracted a great deal of attention as one of the most promising next generation nonvolatile memory devices because of its useful properties such as a simple structure, low cost, and high density integration [1-4]. Because of these merits, various materials such as oxide-based materials [5-7] and nitride-based materials [8-11] have been investigated for their resistive switching (RS) characteristics. However, because the RS materials are not yet optimized for commercial applications of ReRAM devices, until now, various kinds of RS materials have been explored to improve their resistive switching characteristics and reliability. Among them, hafnium oxide ($HfO_2$) is suggested as one of the most promising candidates for ReRAM devices, due to its excellent resistive performance and compatibility with Silicon based semiconductor fabrication processes [12,13]. However, the $HfO_2$ films deposited by physical deposition method mainly investigated by many research groups for improving the RS performance have faced the limitations in the nanoscale fabrication process and low throughput manufacturing level of ReRAM devices. Therefore, in order to overcome these problems, the research on the RS characteristics of $HfO_2$ films grown by chemical deposition is necessary.

In this work, we investigate the $HfO_2$ films grown by metal organic chemical vapor deposition (MOCVD) for the application of ReRAM devices. In addition, to study the resistive switching phenomena of $HfO_2$ films, we analyzed the properties of $HfO_2$ films and investigated the current–voltage ($I$–$V$) characteristics and reliability characteristics of memory cells.

## II. EXPERIMENTS AND DISCUSSION

So as to prepare the samples, a Ti adhesion layer was deposited on the $SiO_2$/Si substrate, after which a TiN bottom electrode of 20-nm thickness was deposited by using a radio frequency sputtering system. Subsequently, $HfO_2$ films with 25-nm was deposited by MOCVD at a substrate temperature of 400 °C. Next, we deposited a 200-nm thick Pt top electrode with a 50 µm diameter using an electron-beam evaporation system. Finally, the sample was annealed in nitrogen ambient at 400 °C for 30 min. Fig. 1(a) shows the schematic drawings of the fabricated $HfO_2$ based memory cells with simple metal insulator metal (MIM) structures, and the measurement configuration. We measured the electrical properties of memory cells using a Keithley 4200 semiconductor parameter analyzer.

First, in order to examine the structural property of the $HfO_2$ films, the surface of $HfO_2$ films was investigated by using atomic force microscopy (AFM) in a non-contact mode having the scan rate of 0.4 Hz. As shown in Fig. 1(b), we observed a number of small $HfO_2$ grain (i.e., amorphous structures) and in this figure, the height between lowest and highest was <1 nm when evaluating line profile in the whole surface. Then, the roughness of $HfO_2$ films with a scanning range of 1000×1000 nm and the difference between the highest point and the lowest point is <0.3 nm. These results mean that the deposited $HfO_2$ films can be uniformly formed by employing the MOCVD method to deposit it as active layer, compared to physical vapor deposition.

In addition, we also measured the x-ray diffraction (XRD) spectra in order to examine the structural properties of $HfO_2$ films with the angular range from 10° to 80°. To analyze the measured results, we used Joint Committee on Power Diffraction Standards (JCPDS) such as, #34-104 for the $HfO_2$ films and #38-1420 for TiN films. As shown in Fig. 1(c), there are broad diffraction peaks at near $32^o$ of (111), $52^o$ of (202), and $62^o$ of (122), related to the $HfO_2$ films, which belong to the amorphous structure. In this formulation, the atoms are arranged in random fashion with no order.

In order to confirm the RS characteristics, we investigated the electrical properties of $HfO_2$ films. Figure 2 (a) shows a typical *I–V* characteristic measured from the $Pt/HfO_2/TiN$ memory cell with a 25-nm thick $HfO_2$ layer, where bipolar resistive-switching behavior is observed by sweeping the voltages in a sequence of 0 V → +6.5 V → 0 V → -3 V → 0 V at room temperature. In order to obtain the first set operation, an initial forming process of 6 V is required because of its own high resistance; conducting filaments can be formed by connecting oxygen vacancies between top and bottom electrodes in forming process, as shown in Fig. 2(b). After that, by dc-sweeping the voltage from 0 V to

+6.5 V, the current level abruptly increased and reached the low resistance state (LRS) with set operation, and the cell remained at the LRS during sweeping back to 0 V and below reset voltage (<$V_{RESET}$). Subsequently, the current of the cell abruptly decreased at a negative voltage of -1.5 V ($V_{RESET}$) under sweeping from 0 V to -3 V, which indicates that the cells change to the high resistance state (HRS) with the reset operation. On the other hand, since we used materials with different work function of top- and bottom-electrodes, we could find a different current level in both bias polarity.

Then, in order to examine the RS performance as well as reliability of Pt/HfO$_2$/TiN memory cells, we investigated the dc cycling characteristics as shown in Fig. 3(a). During 40 repetitive positive and negative bias sweeping operations, they show stable RS characteristics maintaining the stable form of *I-V* curves. On the other hand, as increasing the number of dc cycling, the $V_{RESET}$ has been increased while the set voltage ($V_{SET}$) has been decreased, which is because of the defects induced during the degradation of the HfO$_2$ films under repetitive biases. During the set process, the conducting filaments in HfO$_2$ films can be more easily formed through these defects. On the other hand, during the reset process, it is more difficult to rupture the conducting filaments as increasing the number and size of conducting filaments by too much defects, and it means the difficulties in control of conducting filaments of HfO$_2$ films. Figure 3(b) shows the enlarged *I-V* curves under positive bias for a closer look of increased current level under the repetitive dc cycling test.

Finally, we investigated the distribution of $V_{SET/RESET}$ and currents at the LRS and HRS, and the retention characteristics in order to examine the reliability of Pt/HfO$_2$/TiN memory cells. As shown in Fig. 4(a), they show stable RS performance with small variations of the voltages and currents before and after programming. Especially, a large voltage window of >4.5 V between $V_{HRS}$ and $V_{LRS}$ was obtained, which is enough to distinguish both operating voltages. In this cell, both the HRS and LRS followed the conducting filament model consisted of oxygen vacancies. We speculate that the basic switching mechanism of HfO$_2$-based ReRAM might be closely related to the Hf$^{n+}$, O$^{n-}$ and electron migrations in oxygen-related vacancies. When applying the positive bias over $V_{SET}$, the conduction path can be formed across the bottom- and top-electrodes via set process. Conversely, when applying negative bias over $V_{RESET}$, the conduction path can be ruptured via the ionization of conducting filaments in reset process. The retention time for the LRS, on the other hand, is determined by the thermal release time, which is exponentially proportional to $\Delta E_t$ as $\tau \propto \exp(\Delta E_t/kT)$. Accordingly, a long retention time is expected from the materials having high activation energy ($E_t$). In the retention test, we could not observe the failure phenomena of both the LRS and HRS. In other words, good retention characteristics have been identified without any degradation in both LRS and HRS over $10^5$ s for Pt/HfO$_2$/TiN memory cells, shown by the double logarithmic plot in Fig. 4(b).

## III. CONCLUSION

In conclusion, the HfO$_2$ films grown by metalorganic chemical vapor deposition (MOCVD) for the application of ReRAM devices were investigated. As a result, we observed the bipolar resistive switching characteristics in the 25-nm thickness of HfO$_2$ films with the amorphous structure. In the fabricated Pt/HfO$_2$/TiN memory cells, the stable RS performance was observed under 40 repetitive dc cycling test with the small variations of set/reset voltages and currents before and after programming. In addition, the good retention characteristics over $10^5$ s observed in this cell indicate that the HfO$_2$ films grown by MOCVD are promising candidates for RS memory applications.

## ACKNOWLEDGEMENT

H.-D. Kim and M.J. Yun contributed equally to this work. This research was supported by Basic Science Research Program through the National Research Foundation of Korea (NRF) funded by the Ministry of Education (No. 2015R1D1A1A01056803).

Figure Captions.

Fig. 1. (a) Schematic drawings of the Pt/HfO$_2$/TiN memory cells and the measurement system. (b) Typical AFM topography for the roughness of HfO$_2$ films (c) X-ray diffraction curves measured for HfO$_2$ films.

Fig. 2. (a) Typical current–voltage (*I*–*V*) curve characteristics at room temperature for Pt/HfO$_2$/TiN memory cells. (b) Forming process for the initial set operation.

Fig. 3. (a) DC cycling characteristics during 40 times. (b) Enlarged *I-V* curves during dc cycling test in the positive polarity.

Fig. 4. (a) Probability plots of the operating voltage and current at $V_{READ}$=1 V in LRS and HRS. (b) Retention characteristics of the Pt/HfO$_2$/TiN memory cells.

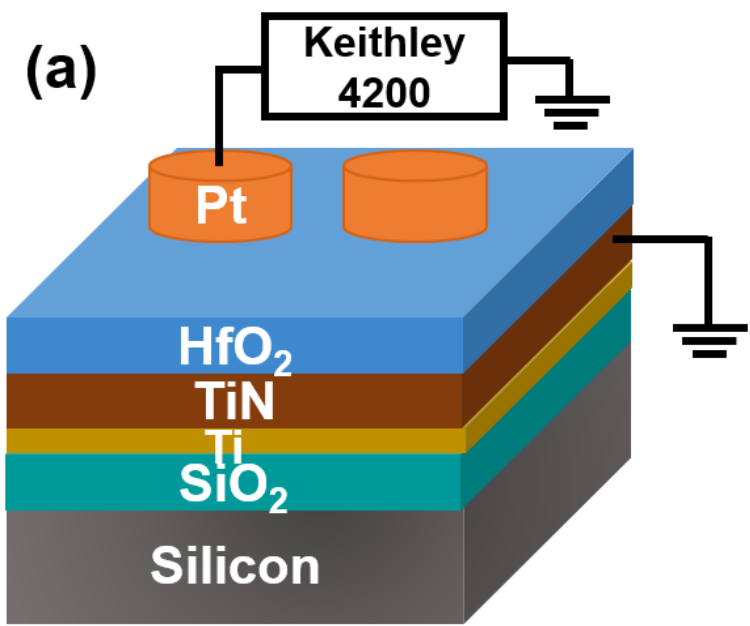

Figure 1(a)

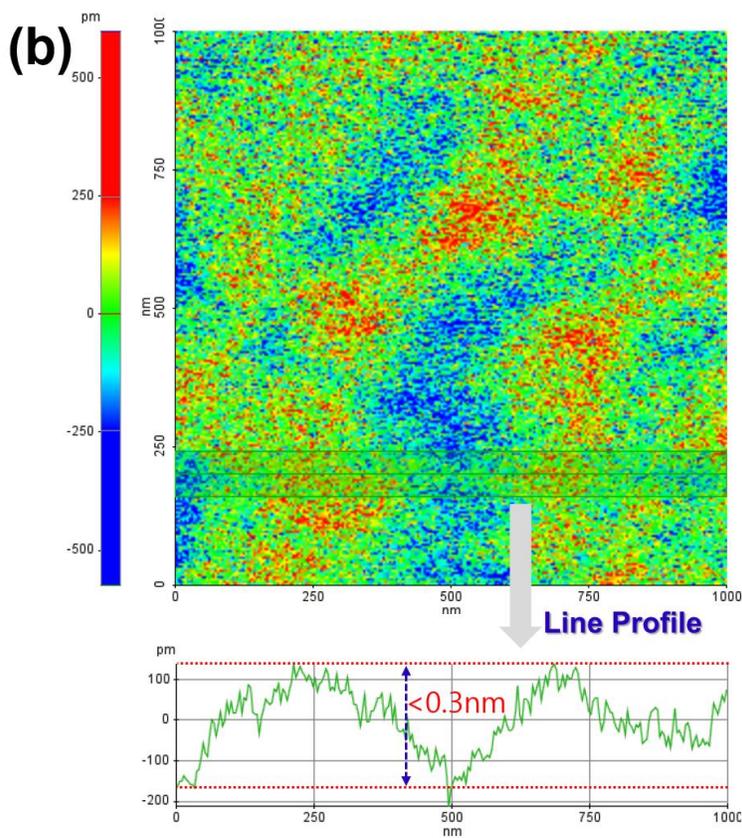

Figure 1(b)

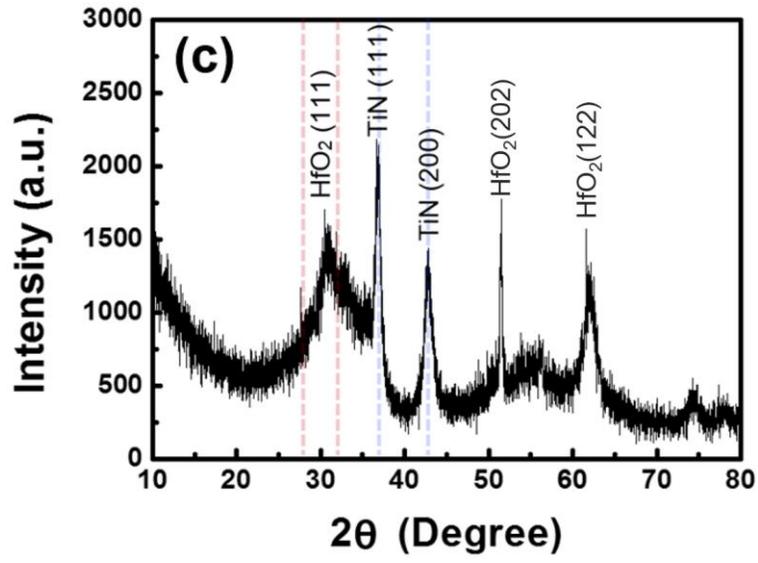

Figure 1(c)

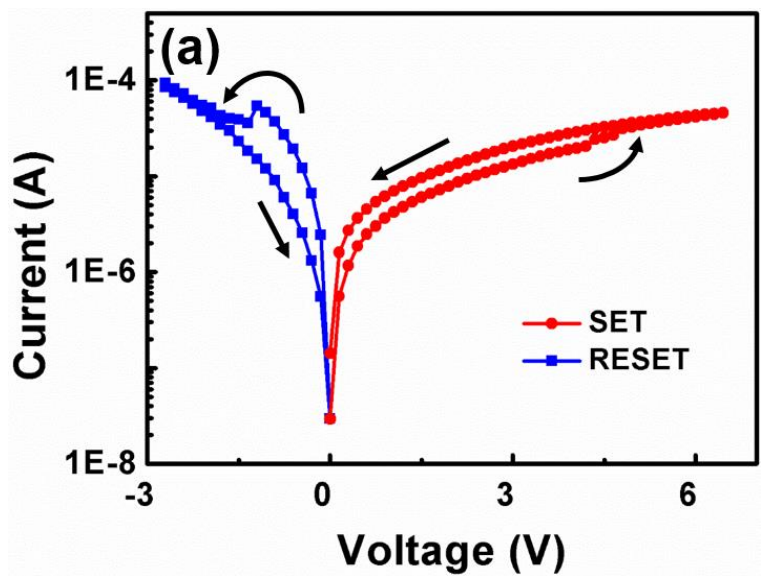

Figure 2(a)

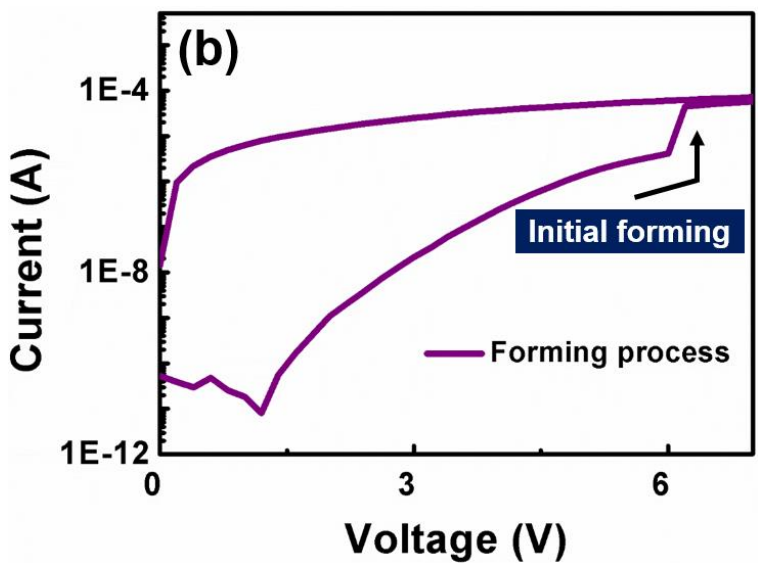

Figure 2(b)

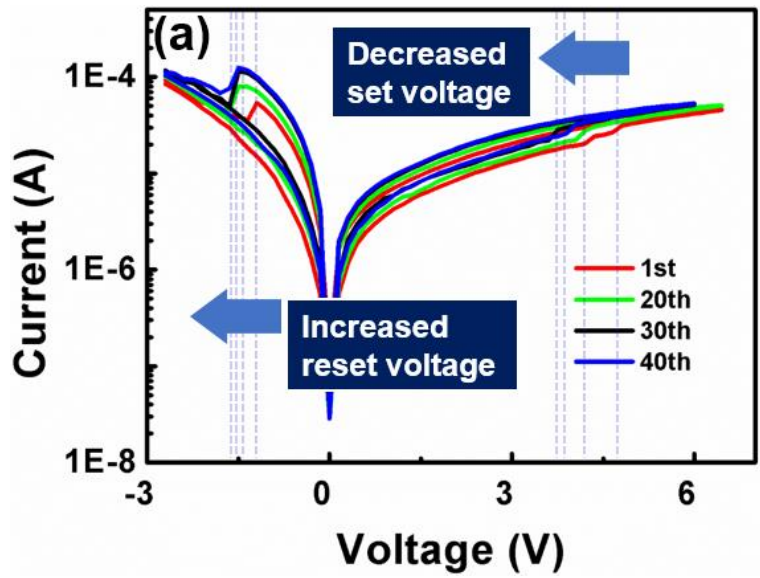

Figure 3(a)

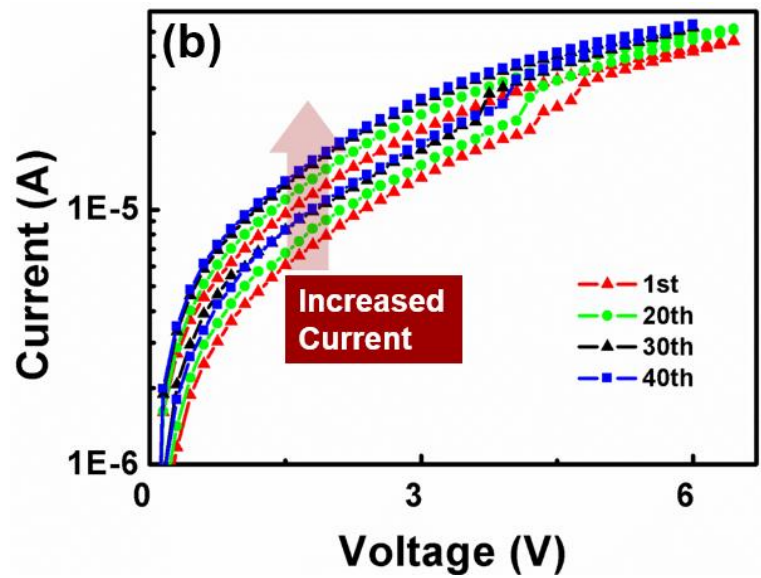

Figure 3(b)

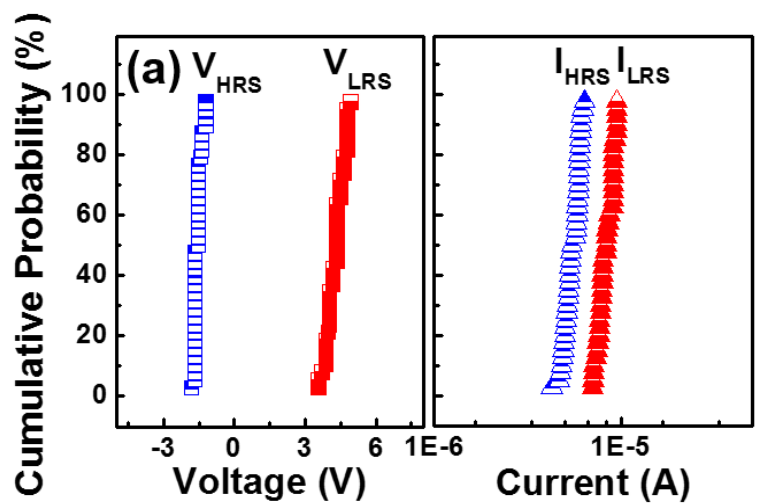

Figure 4(a)

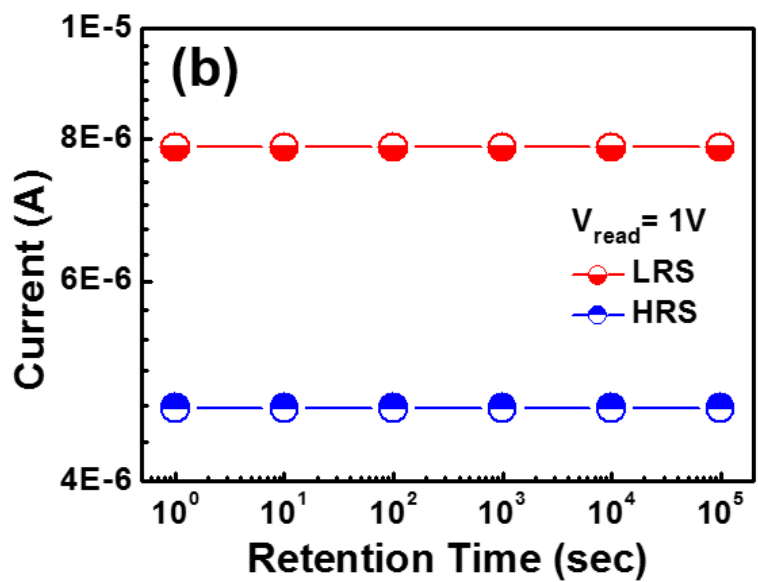

Figure 4(d)